\documentclass[12pt]{article}

\usepackage{cite}
\usepackage{url}
\usepackage{epsfig,endnotes,amsmath}
\usepackage{graphics}
\usepackage{epsfig}
\usepackage{color}

\begin{document}

\title{Neighbor-Specific BGP: More Flexible Routing Policies While Improving Global Stability}
\author{
  Yi Wang, Michael Schapira, Jennifer Rexford\\
  \{yiwang, jrex\}@cs.princeton.edu, michael.schapira@yale.edu}
\date{}
\maketitle

\noindent
\emph{Please Note:}
This document was written to summarize and facilitate discussion regarding
(1) the benefits of changing the way BGP selects routes to
selecting the most preferred route allowed by export policies, or more generally, 
to selecting BGP routes on a per-neighbor basis,
(2) the safety condition that guarantees global routing stability under the 
\emph{Neighbor-Specific BGP} model, and
(3) ways of deploying this model in practice. 
A paper presenting the formal model and proof of the stability conditions was
published at SIGMETRICS 2009 and is available online~\cite{nsbgp:sigmetrics09_link}.

\section{Introduction}
\label{sec:introduction} 

BGP is used by tens of thousands of independently operated networks (i.e., ASes) in the 
Internet to exchange reachability information. However, in current BGP (and ever since it 
was first introduced almost twenty years ago~\cite{rfc1267}), 
a router is restricted to selecting a single best route (for each destination), and either 
exporting this route or no route to each neighbor. 
This restriction has two adverse implications:
\begin{itemize}
\item {\bf Correctness problems:}
A router may not announce any route to a neighbor even if a valid route is available, 
which could cause an ISP to violate its agreements with its neighbors (i.e., policy violation), 
or lead to routing oscillation.
\item {\bf Lack of customizable route selection:}
An ISP cannot capitalize its path diversity by offering customized route selection services
to different neighbors.
\end{itemize}

We discuss the two issues in Section~\ref{sec:correctness} and~\ref{sec:ns-bgp}, respectively.
In each section, we first motivate the problem using examples, then present the corresponding 
solution and discuss how the solution can be deployed in practice. 
To address the correctness problems, we propose to apply export filtering policies
\emph{before} the route selection process; to provide customizable route selection, we propose
\emph{Neighbor-Specific BGP}, which selects routes on a per-neighbor basis.
By using existing mechanisms like Virtual Routing and Forwarding (VRF), encapsulation,
BGP add-paths, etc., both solutions can be incrementally deployed by a single ISP with only 
software changes to its routers. 
\section{Fixing BGP's Correctness Problems}
\label{sec:correctness} 

We illustrate two correctness problems of BGP by examples, and then propose
a simple solution to both problems and discuss how it can be deployed 
by individual ISPs incrementally.

\subsection{BGP's Correctness Problems}
\label{subsec:correctness_problem}

An ISP usually has multiple policy objectives it wants to realize through its BGP
configuration, such as ``consistent export"~\footnote{That is, an AS must make each destination
reachable at every peering point with a neighbor via ``equally goodÓ routes~\cite{Feamster2004}.},
``hot-potato routing"~\footnote{Or, ``early exit" routing, a router selects the ``closest" egress
point in terms of the intradomain path costs, in order to reduce the network
resources required to carry the traffic~\cite{Feamster2004}.},
``provide no transit service for peers or providers", etc.
However, even these natural and appealing policy objectives can be in conflict under
the current BGP, as illustrated in the examples below.

\subsubsection{Problem 1: Policy violation}

We first give an example in which the way BGP selects and exports route
leads to inevitable violation of its export agreement with its neighbor.

\begin{figure}[h]
 \centering
 \includegraphics[width=0.5\textwidth]{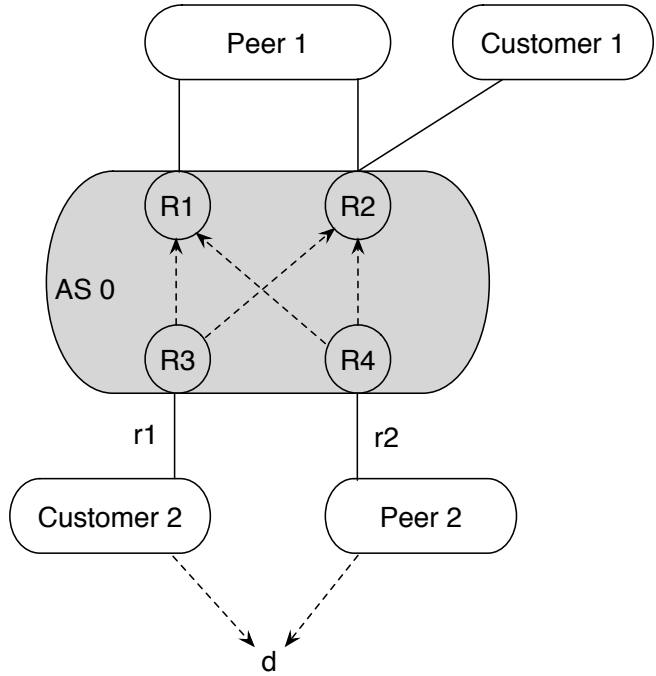}
 \caption{In conventional BGP, R2 has to either export r1 to both Peer 1 and Customer 1, 
 thus violating its hot-potato policy of choosing the closest exit R4 (i.e., r2); or it has to export r2
 to Customer 1 but export nothing to Peer 1 (as r2 is a peer-learned route), thus violating
 the consistent export policy with Peer 1.}
\label{fig:violation}
\end{figure}

Neighboring ISPs ``peer" with each other to reach each other's customers.
It is common practice for ISPs to require ``consistent export" in their peering contracts
when they are connected in multiple geographic locations~\cite{Feamster2004}.
Consider the topology in Figure~\ref{fig:violation}. AS 0 learns
two routes (r1 and r2) to destination d from Customer 2 and Peer 2 respectively.
Assume AS 0 equally prefers customer- and peer-learned routes, and r1 and r2 
are equally preferred in the first few steps of the BGP
decision process (e.g., local preference, AS-path length, etc.).
Due to hot-potato routing, router R1 selects r1 and router r2 selects
r2. In this case, R1 will export r1 to Peer 1 whereas R2 will filter r2
and export no route to Peer 1, because r2 is a peer-learned route
and AS 0 does not provide transit service for its peers.
This results in a violation of the ``consistent export" policy,
which requires AS 0 to export ``equally good" routes through all peering links.
AS 0 could satisfy the consistent export policy by having R2 select r1 as 
the best route, so that Peer 1 learns r1 from both R1 and R2. 
However, Customer 1 will be ``forced" to use r1 as well, even if AS 0
wants to do hot potato routing and use r2 for Customer 1's traffic.

In summary, the one-route-fits-all restriction of the BGP route
selection process makes it impossible for AS 0 to export routes consistently
for Peer 1 and implement hot-potato routing for Customer 1's traffic
at the same time, even though both policies are completely natural
and reasonable.\\

\subsubsection{Problem 2: Routing Oscillation}

\begin{figure}
 \centering
 \includegraphics[width=0.5\textwidth]{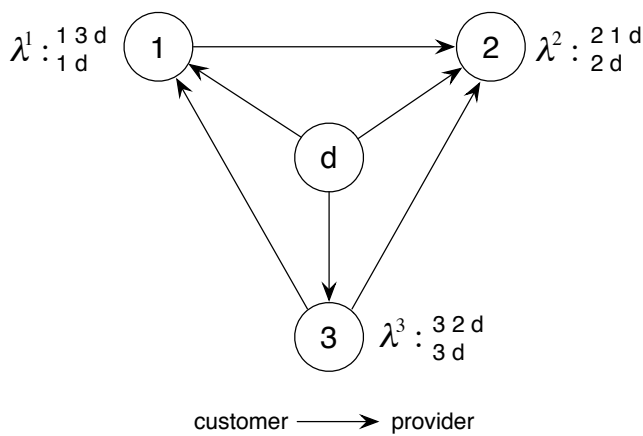}
 \caption{In this BGP ``Bad Gadget", the routing system will oscillate forever. (Each
 node i's preference of routes is specified in the ranking function $\lambda^i$.
 Each node picks the highest-ranked route that is consistent with its neighbors' choices.)}
\label{fig:bad_gadget}
\end{figure}

Figure~\ref{fig:bad_gadget} shows a routing
system in which BGP will always diverge, which is called BGP
\texttt{Bad Gadget}~\cite{Griffin2002c}. 
In this example, $\lambda^1$, $\lambda^2$ and
$\lambda^3$ are the \emph{ranking functions} used by nodes 1, 2 and 3
to select the best route, respectively. It is easy to see that 
BGP will never converge in this system. For example, the routes
chosen by nodes 1, 2 and 3 could change over time in the following
sequence: ($(1\ d)$, $(2\ d)$, $(3\ d)$)
$\rightarrow$ \fbox{($\underline{(1\ 3\ d)}$, $(2\ d)$, $(3\ d)$)}
$\rightarrow$ ((1\ 3\ d), $(2\ d)$, $\underline{(3\ 2\ d)}$)
$\rightarrow$ ($\underline{(1\ d)}$, $(2\ d)$, $(3\ 2\ d)$)
$\rightarrow$ ($(1\ d)$, $\underline{(2\ 1\ d)}$, $(3\ 2\ d)$)
$\rightarrow$ ($(1\ d)$, $(2\ 1\ d)$, $\underline{(3\ d)}$)
$\rightarrow$ ($\underline{(1\ 3\ d)}$, $(2\ 1\ d)$, $(3\ d)$)
$\rightarrow$ \fbox{($(1\ 3\ d)$, $\underline{(2\ d)}$, $(3\ d)$)}.
(An underlined route indicates that it has changed from the
previous state.) Notice that the second state of the system is
the same as the last one. Therefore, the system
will continue to oscillate and never terminate.

\subsection{A Simple Solution to Both Problems}
\label{subsec:correctness_fix}

Both the policy violation and the oscillation problems of BGP stem from the same root cause:
a router chooses the single best route (for all neighbors) \emph{before} applying export policies.
As a result, \emph{a router may not export any route to a neighbor because of the presence or 
absence of another route}.

In the policy violation example (Figure~\ref{fig:violation}), r1 is not exported to Peer 1
because of the presence of r2: R2 selects r2 as the best route (for all of its neighbors),
but as a peer-learned route, r2 is filtered by the ``no transit service for peers" export policy. 
If r2 was not available, R2 would have chosen r1 instead, and Peer 1 would have gotten
consistent export from AS 0. 
In the route oscillation example (Figure~\ref{fig:bad_gadget}), when node 3 learns the route
$(2\ d)$ from node 2, it switches its best route from $(3\ d)$ to the more preferred $(3\ 2\ d)$,
and withdraws route $(3\ d)$ from node 1.
However, since $(2\ d)$ is a provider-learned route and node 1 is another provider of node 3,
$(2\ d)$ is filtered by the ``no transit service for providers" export policy. This results in a 
\emph{pathological} situation: node 3 is exporting \emph{nothing} to node 1, 
even though the previously announced route $(3\ d)$ is still available. 
As illustrated by the oscillation sequence earlier, the withdrawal of route $(3\ d)$ 
by node 3 is the direct cause of the route oscillation, which is due to the presence (availability) 
of route $(2\ d)$.

Both correctness problems can be fixed by simply applying export policies \emph{before} the 
BGP route selection
process. That is, \textbf{only select the best route among the available \emph{exportable} routes 
for each neighbor}. In Figure~\ref{fig:violation}'s example, after applying the export policy
for Peer 1, the only exportable route left to choose from is r1, so R2 would export
r1 to Peer 1. However, both r1 and r2 are exportable for Customer 1, so R2 can choose
the one most preferred by its ranking function as the best route (i.e., in this case, r2),
and export it to Customer 1. This way, AS 0 can simultaneously satisfy the consistent export 
policy with Peer 1 and its own hot-potato policy of forwarding Customer 1's traffic.
Similarly, the oscillation problem can be fixed by only selecting the 
best route among the available \emph{exportable} routes for each neighbor. In this example,
although node 3 learns two routes to reach d (i.e., $(3\ d)$ and $(3\ 2\ d)$), only $(3\ d)$ is
exportable to node 1, so node 3 would export route $(3\ d)$ to node 1. At the same time,
node 3 could choose route $(3\ 2\ d)$ to forward its own traffic to d. If node 1 and 2 also
follow the same principle, the system will converge to a stable state where every node
gets its most preferred route.

\subsection{Deployment}
The above solution to BGP's correctness problems can be deployed by simply
moving the ``export filtering" step before the route selection process. In practice, 
since export filtering is usually performed according to business relationships,
which, in turn, are determined by neighbor types,
grouping export policies into several classes according to neighbor types
(such as ``customers", ``peers" and ``providers") can achieve the benefit with
little extra configuration overhead.
Since this modification could result in different best routes for different neighboring
domains, an iBGP speaker may need to disseminate extra routes besides its own best 
route. However, the number of extra routes that need to be disseminated is not
large, e.g., at most 2x in the case where there are two main classes of export
policies --- ``export all routes" and ``export only customer-learned routes".
Existing MPLS or IP-in-IP tunneling mechanisms can be used to ensure that traffic 
from different ingress points is forwarded to the correct egress points, as detailed
in Section~\ref{subsec:ns-bgp_system}.

\section{Making BGP Route Selection Customizable}
\label{sec:ns-bgp}

\subsection{BGP's Lack of Customized Route Selection}
\label{subsec:ns-bgp_case}

Besides causing correctness problems, 
BGP's one-route-fits-all restriction also makes it impossible for an ISP to
capitalize its path diversity by offering customized route selection services.
Different neighbors of an ISP may have completely different
preferences for the kinds of paths that should carry their traffic.
For example, an online gaming provider may prefer paths with low
latency, whereas a financial institution may prioritize security over
performance.  However, in today's BGP,
each router selects and advertises a \emph{single} best route,
limiting an AS's ability to offer customized route selection for its
neighbors.  

We argue that an extension to BGP that allows a router to offer different 
interdomain routes to different neighbors is beneficial to ISPs
for three main reasons:

{\bf (1) Many ISPs have rich path diversity.} 
ISPs offering transit service usually connect to many other ASes,
often at multiple locations~\cite{wolfgang:sigcomm06,ratul:nsdi07}.
As a result, it is quite common for large networks to have 5-10 paths
per prefix, with some prefixes having more than 20 different
paths~\cite{nanoglist:diversity}.

{\bf (2) Different paths have different properties.}
The many alternative paths an ISP has could have different
security~\cite{pgbgp:cn} and performance~\cite{savage:sigcomm99} properties. 
In fact, alternative interdomain paths often have significantly better 
performance than the paths chosen by BGP~\cite{kobus:pam07}.

{\bf (3) Different neighbors may want different paths.}
Different neighbors of an ISP may have very different preferences
on the types of paths they get from the ISP. For example, financial 
institutions may prefer the most secure paths (e.g., paths that avoid
traversing untrusted ASes, such as ASes known to censor traffic), 
while providers of interactive applications like online gaming and VoIP may
prefer paths with low latency. If such options were available, they
might be willing to pay a higher price to have the paths they want. 
Yet some other neighbors may be perfectly happy with whatever
paths the ISP provides for a relatively low price.

\begin{figure}[h]
 \centering
 \includegraphics[width=0.5\textwidth]{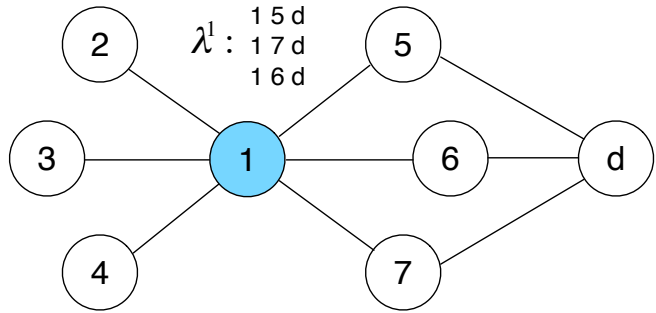}
 \caption{In conventional BGP, node 1 uses a single ranking function ($\lambda^1$)
 to select routes for all neighbors, making it impossible for nodes 2, 3, and 4 to 
 use the different routes available to reach the same destination d.}
\label{fig:conventional_bgp}
\end{figure}

Ideally, an ISP should be able to offer different routes to different
neighbors, regardless of whether they connect to the same edge router.
However, this is not possible with conventional BGP. For example,
in Figure~\ref{fig:conventional_bgp}, even though neighbor 2, 3, and 4
would like to use the three different routes node 1 has to reach d,
node 1 uses a single ``ranking function" to select the best route,
forcing the three neighbors to use the same route.

\subsection{Customized Route Selection with Neighbor-Specific BGP}

In~\cite{nsbgp:sigmetrics09_link}, we proposed \emph{Neighbor-Specific BGP}
(NS-BGP), a simple extension to BGP that allows a router to select routes on a 
per-neighbor basis.
NS-BGP inherits everything from conventional BGP (from the message 
format to the way messages are disseminated between ASes) except 
for how it selects routes. As a result, an individual ISP can
independently deploy NS-BGP and offer value-added route-selection 
services.  All the changes required for an AS to deploy NS-BGP are 
within its own network and practically feasible, as discussed in
Section~\ref{subsec:ns-bgp_system}. 

\begin{figure}
 \centering
 \includegraphics[width=0.6\textwidth]{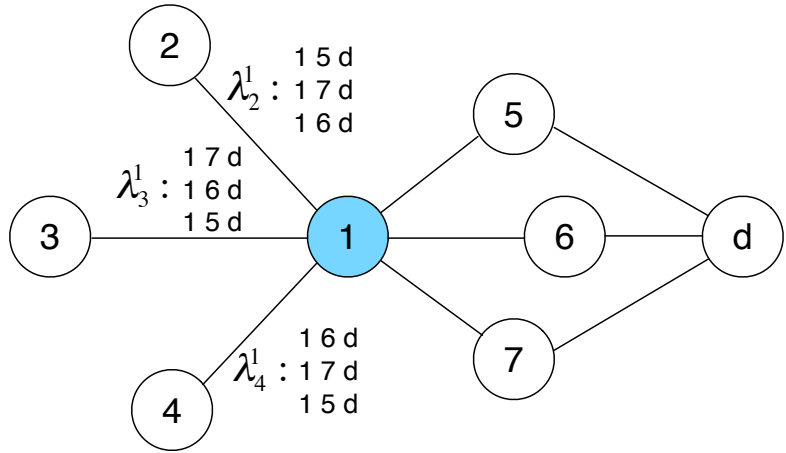}
 \caption{In NS-BGP, node 1 uses three different ranking functions ($\lambda_2^1$,
 $\lambda_3^1$, and $\lambda_4^1$) for the three neighbors 2, 3, and 4, allowing
 them to use different routes to reach the same destination d.}
\label{fig:ns_bgp}
\end{figure}

Recall that in conventional BGP, a router uses a single ``ranking function"
to select best routes for all its neighbors. Even in the fix to the BGP
correctness issues described in Section~\ref{subsec:correctness_fix}, a router still
uses a single ranking function (but the same ranking function is used
to select the best route from potentially different sets of exportable routes
for different neighbors).
In NS-BGP, however, a router {\bf can use multiple \emph{different} ranking functions
to select routes for different neighbors}. With this additional flexibility,
a router can select a different best route for different neighbors even with
the same set of available routes.
For example, in Figure~\ref{fig:ns_bgp}, node 1 uses a different ranking
function for each neighbor, therefore allowing nodes 2, 3, and 4 all have
their most preferred route.

\subsection{Deployment}
\label{subsec:ns-bgp_system}
With the availability of Virtual Routing and Forwarding (VRF), MPLS/IP-in-IP 
tunneling mechanisms and the on-going development of BGP add-paths
capability~\cite{bgp:add-path}, NS-BGP is practically feasible and 
can be deployed by individual ISPs independently.
We first describe how an AS can correctly
forward traffic from different neighbors (and from within its own 
network) along different paths. We then discuss how to 
disseminate multiple routes to the edge routers of an AS to enable
flexible route selection. Finally, we present three models an 
NS-BGP-enabled AS can use to provide different levels of 
customized route-selection services.
When deploying NS-BGP, an AS can handle all these issues
by itself without requiring any changes from neighboring ASes,
as no BGP message format or external BGP (eBGP) configuration
changes are needed.

\subsubsection{Neighbor-Specific Forwarding}
NS-BGP requires routers to be able to forward traffic from different 
neighbors along different paths. Fortunately, today's routers 
already provide such capabilities. For example, the ``virtual routing 
and forwarding (VRF)" feature commonly used for Multi-protocol 
Label Switching Virtual Private Networks (MPLS-VPNs) supports
the installation of different forwarding-table entries for 
different neighbors~\cite{mpls:vpn}. 

Since an AS typically consists of many routers, traffic
entering from various \emph{ingress} routers of the AS must be 
forwarded to the correct \emph{egress} routers. In conventional BGP,
this is achieved in a hop-by-hop fashion to ensure that all routers in 
the AS agree to forward traffic to the closest egress 
point that has one of potentially multiple ``equally good" best
paths to the destination. For example, in Figure~\ref{fig:principles}, 
if $R5$ learns one path from $R3$ and another path from $R4$ to 
$D$, and the two routes are considered ``equally good" in BGP's 
route-selection process, it will choose to use the closest egress 
point (according to the IGP distances).
However, this approach no longer works in NS-BGP,
as traffic entering the AS at the same ingress point may be from
different neighbors (ingress links), and thus may need to be forwarded
to different egress points, or different egress links of the same
egress point. Fortunately, ASes have an efficient solution
available---encapsulation (or tunneling). Many commercial routers
deployed in today's networks can perform MPLS or IP-in-IP 
encapsulation / decapsulation at line rate. To provide customized
forwarding for neighbors connected at the same edge router,
the tunnels need to be configured from ingress \emph{links} (rather than
ingress routers) to egress \emph{links} (rather than egress routers).
For example, in Figure~\ref{fig:principles}, $C1$'s and $C2$'s traffic 
can be tunneled from $R1$ to $R6$ and $R7$ 
(that connect to the same egress point $R3$) independently.
To avoid routers in neighboring domains having to decapsulate 
packets, egress routers need to remove the encapsulation header 
before sending the packets to the next-hop router, using technique 
similar to the penultimate hop popping~\cite{cisco:mpls}.
Similar to transit traffic originated from other ASes, traffic 
originated within the AS itself can also be forwarded to the correct
egress links using tunneling.

\subsubsection{Route Dissemination Within an AS}

\begin{figure}
 \centering
 \includegraphics[width=0.6\textwidth]{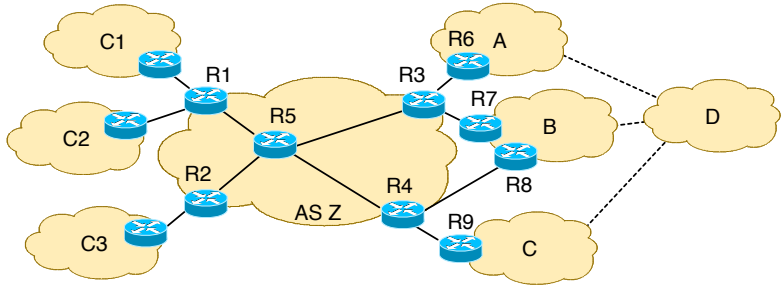}
 \caption{AS Z has multiple interdomain routes for destination D}
 \vspace{-3mm}
\label{fig:principles}
\end{figure}

A prerequisite for an edge router to provide meaningful ``customized"
route-selection services is that it needs to have multiple
available routes to choose from (otherwise, all its neighbors would inevitably
receive the same route). Unfortunately, the way BGP 
routes are disseminated within today's ASes makes such ``route visibility" 
often impossible. For example, in Figure~\ref{fig:principles}, the AS $Z$
as a whole learns four routes to D from four different neighboring
edge routers ($R6$, $R7$, $R8$, $R9$).
However, as BGP only allows a router to select and announce a 
single route for a destination, router
$R5$ will only learn two of the available routes, one from $R3$ and $R4$.
Even worse, $R1$ and $R2$ will only learn the one route selected by $R5$.
For similar reasons, in large ASes where route reflectors are commonly 
used for better scalability, most edge routers have significantly reduced
visibility of BGP routes~\cite{uhlig:networking06}.

To provide better route visibility to the edge routers of an AS, a router in the AS 
needs to be able to \emph{disseminate} multiple routes (per destination)
to each neighbor. For backwards compatibility, this can be achieved 
by using multiple internal BGP (iBGP) sessions between routers. 
The BGP ADD-PATH extension, which supports the dissemination of multiple
routes (per destination) through one BGP session~\cite{bgp:add-path},
makes the dissemination process much more efficient. 
We note that, depending on how much flexibility an AS plans
to provide, \emph{not} all available routes need to be disseminated.
For example, if an AS decides to have a couple of notions of ``best routes"
(e.g., best of all routes, and best of customer-learned routes), it only needs
to disseminate at most two routes per destination (one of which must
be a customer-learned route). 
Different ASes can make different trade-offs between the overhead of 
disseminating more routes within their own networks and the benefit 
of providing more routes to their neighbors to choose from. 

Alternatively, an AS can also improve its route visibility by using a 
logically-centralized Routing Control Platform (RCP)~\cite{nsdi05:rcp, irscp:usenix07, 
morpheus:jsac}. In this case, an AS can deploy a set of servers in its
network, each of which has a complete view of all available BGP routes. 
These servers then select routes on behalf of all the edge routers and 
install the selected routes to the respective routers. 

\subsubsection{Control Over Customized Selection}

A big motivation of NS-BGP is to enable individual ASes to provide 
customized route-selection services to their neighbors. Therefore, an
NS-BGP-enabled AS needs to take its neighbors' preferences of routes 
into account when selecting routes. Here we describe how an AS can control
the amount of customer influence over its route-selection process, and 
how the customized route selection can be realized.

An AS can use different models to grant a neighbor different levels of 
control over the ranking function it uses for that neighbor. For example, it
could adopt a \emph{``subscription"} model, in which it offers several different 
services (ranking functions) for its neighbors to choose from, such as 
``shortest path", ``most secure", and ``least expensive". A neighbor 
has the flexibility to decide which one to ``subscribe'' to, but does not
have direct influence on how the ranking functions are specified.
Although more flexible than conventional BGP, this model is a still fairly 
restrictive. 

For neighbors that require maximum flexibility in choosing 
their routes, an AS could offer a \emph{``total-control"} model. 
In this model, the AS gives a neighbor direct and complete control
over the ranking function it uses for this neighbor. Therefore,
the neighbor is guaranteed to receive its most preferred routes among all
of all available routes. 

For neighbors that require a level of flexibility that is in between what 
the previous two models offer, an AS could adopt a third, \emph{``hybrid"} 
model. In this model, a neighbor is allowed to specify certain preference
to an AS directly (e.g., avoid paths containing an untrusted AS if
possible). When determining the ranking function for that neighbor,
the AS takes both the neighbor's preference and its own preference into account
(as the ``best route" according to the neighbor's preference may not be the
best for the AS's own economic interest).
Nevertheless, the AS still controls how much influence (``weight") 
the neighbor's preference has on the ranking function.

In~\cite{morpheus:jsac}, we described in detail how these different
models can be implemented by using a new, weighted-sum-based 
route-selection process with an intuitive configuration interface.
When deciding which model(s) to offer, an AS needs to consider the 
\emph{flexibility} required by its neighbors as well as the \emph{scalability} 
of its network, as the three service models impose different resource 
requirements on the provider's network. For example, the 
``subscription" model introduces the least overhead in 
terms of forwarding table size, route dissemination and customized route 
selection (e.g., each edge router or RCP server only 
needs to run a small number of route selection processes). 
On the other hand, the ``total-control" model, while providing the finest
grain of customization, imposes the most demanding requirements on 
system resources and results in the highest cost for the provider. 
Therefore, we expect an AS to only provide such service to a 
small number of neighbors for a relatively high price.
Since the costs of offering the three types of service models are in line
with the degrees of flexibility they offer, we believe that an AS can 
economically benefit from offering any one or more of these models 
with appropriate pricing strategy.

It is worth mentioning that the ``hybrid" and ``total-control" models can
be realized in two different ways. The simpler way is that a neighbor 
tells the AS what ranking function to use, so the AS only needs to 
select and export one route to the neighbor.
The other way is that the AS announces all exportable routes to a neighbor, and 
let the neighbor to select amongst them itself. The latter approach allows the 
neighbor to hide its policy (ranking function) but requires the AS's ability 
to export multiple routes to the neighbor, and the neighbor's ability to directly tunnel
its traffic to the AS's egress links.

\subsection{NS-BGP Is Safe}
Despite the benefits of greater flexibility, enhancements to BGP
should not come at the expense of routing instability.  In fact, even
\emph{without} neighbor-specific route selection, today's BGP can easily
oscillate, depending on the local policies ASes apply in selecting and
exporting routes~\cite{griffin:99,Griffin2002c}.  

Fortunately, we were able to prove that the \emph{more} flexible NS-BGP
requires significantly \emph{less} restrictive conditions to guarantee
routing stability, comparing to conventional BGP~\cite{nsbgp:sigmetrics09_link}.  
Specifically, an AS can freely choose
\emph{any} ``exportable'' path (i.e., a path consistent with the
export condition) for each neighbor without compromising global
stability.  That is, an AS can select \emph{any route} for a customer,
and \emph{any customer-learned route} for a peer or provider.  
Intuitively, this is because in NS-BGP, a route announced to a peer or 
provider is no longer dependent on the presence or absence of any 
\emph{non-exportable} (e.g., peer- or provider-learned) routes chosen 
for customers. This result is less restrictive than the best-known
stability conditions for conventional BGP (known as the ``Gao-Rexford
conditions"~\cite{gao:rexford}), which require an AS to always prefer
a customer-learned route over a peer- or provider-learned route.

\small
\bibliographystyle{abbrv}
\bibliography{paper}

\end{document}